\documentstyle[aps,preprint]{revtex}
\begin{document}
\title{Mapping of mutation-sensitive sites in protein-like chains.}
\author{M. Skorobogatiy$^{1*}$ and G. Tiana$^{2,3}$}
\address{$^1$ Department of Physics, MIT, Cambridge, USA}
\address{$^2$ Department of Physics, DTU, building 307, 2100 Lyngby, Denmark}
\address{$^3$ Dipartimento di Fisica, Universit\`a di Milano,
Via Celoria 16, I-20133 Milano, Italy }

\date{\today}
\maketitle

\smallskip
{\centerline
{Pacs numbers: 87.15.Da, 61.43.-j, 64.60.Cn, 64.60.Kw}
}
\begin{abstract}

 In this work we have studied, with the help of a simple on-lattice
model, the distribution pattern of sites sensitive to point mutations
('hot' sites) in protein-like chains. It has been found that this
pattern depends on the regularity of the matrix that rules the
interaction between different kinds of residues. If the interaction
matrix is dominated by the hydrophobic effect (Miyazawa Jernigan
like matrix), this distribution is very simple - all the 'hot' sites can
be found at the positions with maximum number of closest nearest
neighbors (bulk).

If random or nonlinear corrections are added to such an interaction matrix
the distribution pattern changes. The rising of collective effects allows
the 'hot' sites to be found in places with smaller number of nearest 
neighbors (surface) while the general trend of the 'hot' sites to fall 
into a bulk part of a conformation still holds.    

\end{abstract}

\newpage
\narrowtext

\section{INTRODUCTION}

In this paper we study how the choice of a particular Hamiltonian is
responsible for the distribution pattern of sites sensitive to 
point mutations in a Heteropolymeric chain.  

As shown in \cite{my} using
a very simple model \cite{prl94}, in every optimized \cite{eng} sequence
there are sites at which point mutations are likely to cause 
misfolding of the native state (we call them 'hot' sites), while there are
other sites at which point mutations have no relevant thermodynamic
effect ('cold' sites), and we call intermediate sites 'warm'. As known 
from both experimental
\cite{real} and theoretical studies \cite{my} usually proteins are made of
few 'hot' sites while the majority of the other sites are 'cold' .
Because of the thermodynamic importance of the 'hot' sites it is of
general interest to investigate the principles guiding their positioning 
along the protein chain. 

\par
The model used in this work describes the polymer as
a chain of beads in a cubic lattice, interacting through the
Hamiltonian
\begin{equation}
H=\frac{1}{2}\sum_{ij}^L B_{ij}\Delta(r_i-r_j),
\end{equation}
where $B_{ij}$ is the contact energy between the two residues
situated on the $i$--th and $j$--th positions, $L$ is the length of
the chain and $\Delta(r_i-r_j)$
has the value $1$ if the $i$--th and $j$--th are nearest neighbors
and zero if not.
\par
In literature, different choices of the matrix $B_{ij}$ have been used.
The results of Ref. \cite{my,prl94} have been found with a matrix 
\cite{mj1}
whose elements are distributed according to a Gaussian, with average
$\overline{B}=0$ and standard deviation $\sigma=0.3$ (in units of 
$kT=0.6\;kcal/mol$). With these matrix elements, for every target structure
it is possible to select sequences whose native
state is unique, stable and kinetically accessible \cite{my2}. 
It has been found that a reliable condition for a site to be 'cold' or 'hot'
is intimately connected to the change in the native state energy caused by
a point mutation: the bigger is this difference, the more 
probable for this site to be a 'hot' site.  
Hot sites of such sequences are mainly in the bulk sites of the native
conformation, but can be on the surface as well, while some bulk sites
can be rather insensitive to mutations (Fig.\ref{fig1}a).

\par
Repeating the same calculations with a random--generated,
Gaussian--distributed set of interaction energies,
we have observed similar distribution patterns for 'hot' and 'cold'
sites (Fig.\ref{fig1}b).
\par

Another choice of the interaction matrix can be made to take into account
explicitly the hydrophobic effect encountered in real proteins.
The simplest way is to choose the matrix $B_{ij}$ to be
composed of only three different 
elements, namely $B_{HH}$, $B_{PP}$ and $B_{HP}=B_{PH}$, where
$B_{HH}<B_{PP}<B_{HP}$. These elements are responsible for the interaction 
between
hydrophobic (H) and polar (P) residues. Unfortunately it was found
\cite{prl94} that for a given target structure and 'HP' interaction
matrix it is difficult to construct optimal sequences for which this
target structure would be kinetically accessible.  
As a rule, the optimization of the sequence puts H--residues in the
bulk sites (the sites with the greatest number of nearest--neighbors),
and every substitution of an H residue with a P residue causes the
chain misfolding (See Figs. 2 and 3 in Ref. \cite{wingreen}).
\par

On the way between the random matrix and the highly regular HP
interaction model stands the matrix deduces by Miyazawa and Jernigan in
\cite{mj2}. This matrix contains $210$ different elements which can still
be grouped into $3$ big blocks, according to their hydrophobicity.
In this case it is possible to find sequences for which the
native state is both stabile and kinetically accessible. It was also
found that, as in the situation with only two kinds of
residues, 'hot' sites are invariably the sites with the highest 
number of contacts (bulk sites, Fig. \ref{fig1}d).

The question again is what are the principles guiding the positioning 
of 'hot' sites along the protein chain and how a particular form of the
interaction matrix can influence the distribution pattern of 'hot' and 
'cold' mutation sites. This paper is organized as following. In the next 
section we will present a convenient representation of an interaction matrix 
as a function of 'mixing' parameter $\beta$. Variation of the 'mixing' 
parameter will correspond to a change from highly ordered HP like 
interaction matrix at $\beta=0$ to highly nonlinear matrix at non-zero 
values of 'mixing' parameter. A distribution pattern of 'cold' and 
'hot' sites will be investigated with the use of these interaction matrices.
Next we will address the same question by introducing a parameter of 
'randomness' $\gamma$ which would allow us to investigate the 
mutation sites distribution pattern with highly ordered HP like 
interaction matrix at $\gamma=0$ and random Gaussian 
interaction matrix at large $\gamma$. Conclusions will be drawn at the end.

\section{LTW PARAMETERIZATION}
In the work by Lee, Tang and Wingreen \cite{LTW} a
particular interesting parameterization of semi--experimental MJ matrix 
\cite{mj2} was introduced, as a consequence of its regularity. 
In their work it was shown that elements of the MJ interaction matrix 
can be very nicely fitted as 
\begin{equation}
B_{\alpha \gamma}=q_{\alpha}+q_{\gamma}+\beta q_{\alpha}q_{\gamma},
\label{LTW}
\end{equation}
where $q_{\alpha}$ is ascribed to a monomer of kind $\alpha$ and $\beta$
is a constant. This parameterization involves $20$ parameters, instead of 
$210$ 
parameters of the MJ matrix, each specifying the strength of a particular
residue. Using this parameterization we have found that, for the best fit,
all $q_{\alpha}$ are negative and range from $-2.3$ to $0$, while
$\beta=-0.42$.
Furthermore, it has been pointed out that the origin of the additive term
is due to the hydrophobic effect, while the second order term is
responsible for the segregation of dissimilar residues. 

\par
Using this parameterization, we can write the Hamiltonian of the chain in
a form \cite{maksim}
\begin{equation}
H=\vec{q}\vec{n}+\frac{\beta}{2}\vec{q}C\vec{q},
\label{paramH}
\end{equation}
where $\vec{n}$ and $\vec{q}$ have dimensions equal to the number of 
monomers in
the chain. The $i$--th coordinate of $\vec{n}$ is the number of nearest
neighbors for the $i$--th monomer and the $i$--th coordinate of $\vec{q}$ 
specifies the strength of the residue in the
$i$--th site. $C$ is the contact matrix for a given conformation.
In the above reference it has been shown that the Hamiltonian
(\ref{paramH})
fits very well the original MJ Hamiltonian and, what is more
precious, it is very convenient to handle analytically. 
\par
One of the problems that can be solved easily using the above Hamiltonian
is, given a target structure ($C$ matrix and $\vec{n}$) and the composition
of the chain in term of residues, to find the sequence which minimize
the energy \cite{maksim}. Particularly, using the fact that the
second order term is usually $2-3$ times smaller than the first order
term, a first order approximation solution can be found minimizing
only the solvent exclusion term. 
The straightforward way is to choose a sequence so 
that the vectors $\vec{q}$ and $\vec{n}$ are as anti--parallel as possible, 
keeping the constraint of a fixed number of different kinds of monomers.
Knowing that all components of $\vec{q}$ are negative while all components
of $\vec{n}$ are positive, it is necessary to put the residues with the
most negative value of $q_i$ in the sites of the target structure
with the largest number of nearest--neighbors. The effect is,
roughly speaking, to put 'hydrophobic' residues (i.e. low $q_i$) inside
the structure while keeping 'hydrophilic' ones (high $q_i$) on the
surface. The second order term in the Hamiltonian (\ref{paramH}) is
responsible for a fine tuning of residue distribution, mostly 
inside the hydrophobic/hydrophilic regions, and causes a further
decrease of the sequence energy.

\section{MUTATIONS AND HOT SITES}

According to \cite{my}, we label each mutation with the difference
of the native state energy of the wild--type sequence and of the
mutated sequence
\begin{equation}
\Delta E_{loc}=\frac{1}{2}\sum_{ij}(B^0_{ij}-B^\alpha_{ij})
\Delta(r_i-r_j),
\end{equation}
where $B^0_{ij}$ is the interaction element associated with the 
wild--type sequence and $B^\alpha_{ij}$ with a mutated sequence. It has 
been shown that the information about
the thermodynamic features of the mutated sequence is mostly contained 
in the value of $\Delta E_{loc}$ \cite{my2}, \cite{busse}.
\par
The energetic effects of mutations in a given site $i$ are studied
introducing the average $\Delta E_{loc}$ over the $19$ possible mutations
in this site,
\begin{equation}
\overline{\Delta E_{loc}}(i)=\frac{1}{19}\sum_\alpha^{19}\sum_{j}
(B^0_{ij}-B^\alpha_{ij}) \Delta(r_i-r_j).
\end{equation}
For optimized sequences there are few sites (up to $10\%$) where 
$\overline{\Delta E_{loc}}$ is large ($\sim \Delta$, where $\Delta$ is the
gap between the native state and the random conformations energy \cite{prl94}), 
while for the others $\overline{\Delta E_{loc}}$ is much lower. We
call the former sites 'hot' and the latter 'cold'. If a
mutation occurs in a 'hot' site, there is a high probability that 
it fills the gap, eliminating the feature of design and causing
misfolding of the chain.
\par
For random interaction matrices whose elements have Gaussian distribution, the
behavior of hot sites is the same as indicated in Ref. \cite{my}. Hot sites
can be both in the bulk and on the surface of the native configuration  
(See Fig. \ref{fig1}b) with dominance of 'hot' sites in the bulk.
\par
On the other hand, in a model with only two kinds of residues (H and P),
the optimization of the sequence puts H residues in the sites with
the highest number of nearest neighbors. In this case (see 
Ref. \cite{wingreen}), every site containing a H residue is a hot site. 

\par
We are now interested in studying the pattern of hot sites in a model 
where the interaction matrix contains both features of hydrophobic-hydrophilic
separation and randomness or complicated non--linearity.

\section{CONSTRUCTION OF THE MAP OF 'HOT' AND 'COLD' SITES} 

To investigate the behavior of $\overline{\Delta E_{loc}}(i)$ we use
the parameterized Hamiltonian (\ref{paramH}). It is straightforward to show
that, for a wild--type sequence characterized by $\vec{q}$ 
\begin{equation}
\overline{\Delta E_{loc}}(i)=-(n_i+\beta \sum_j^L C_{ij}q_j)(q_i-<q>_{\alpha}),
\label{site}
\end{equation}
where $<q>_{\alpha}=1/20\sum_{\alpha=1}^{20}q_{\alpha}$, is the 
average of the
values $q_{\alpha}$ corresponding to the $20$ monomers different from the 
wild--type. With this assumption we have found $<q>_{\alpha}\approx -1.23$.
\par
We shell consider first the case $\beta=0$, which is exactly solvable, and
then the consequences of the non-linear term introduced via non--zero 
'mixing' parameter $\beta$..

\subsection{$\beta=0$}

In the case $\beta=0$ the Hamiltonian contains only the solvent exclusion 
term $H_0=\sum_i q_i n_i$. As we discussed earlier, optimization of
the sequence, given a structure, consists in putting 
the most hydrophobic residues into
the sites with the greatest number of contacts, keeping with the
constraint of fixed number of different residues, so that
\begin{equation}
q_i\approx\frac{\sum_{i=1}^L q_i}{\sum_{i=1}^L n_i}n_i.
\label{opt}
\end{equation}
\par
For short chains and 20 letter code 
used in the model with $\beta=0$ the optimization procedure described above 
works reasonably good in constructing optimal sequences while for longer chains
it seems that the model is not adequate to ensure a single ground state 
or a ground state well--separated from the other states.
Nevertheless, the case $\beta=0$ is a good starting point to investigate 
the role of the different terms of Hamiltonian (\ref{paramH}).
\par
In this case equation (\ref{site}) can be written as
\begin{equation}
\overline{\Delta E_{loc}}(i)=-n_i(q_i-<q>_{\alpha}).
\label{b0}
\end{equation}
Defining $<q>=\sum_{i=1}^Lq_i / \sum_{i=1}^Ln_i$ and substituting (\ref{opt})
in (\ref{b0}) as an approximation for an optimal sequence, we obtain
\begin{equation}
\overline{\Delta E_{loc}}(i)=-n_i <q> (n_i-\frac{<q>_{\alpha}}{<q>}).
\end{equation}
For typical values of $<q> \sim -0.6$ and $<q>_{\alpha} \sim -1.26$ then
\begin{equation}
\overline{\Delta E_{loc}}(i)\approx 0.6 n_i (n_i-2.1).
\end{equation}
The shape of this function is plotted in Fig. \ref{fig2}.  
It is interesting to compare the value $\overline{\Delta E_{loc}}(i)$ for
the optimized sequence with the one for a random sequence. We will
identify, in the spirit of Random Energy Model \cite{derrida}, 'hot' 
sites (as defined in Sect. III) with those sites in which the average 
impact of
mutations is greater than for a random sequence. We define a sequence
to be random if there is no correlation between the strength $q_i$ at
a given site, and its number of nearest neighbors $n_i$, so that
\begin{equation}
\overline{\Delta E_{loc}^{rand}}(i)=-n_i(q-<q>_{\alpha}),
\end{equation}
where $q$ ranges between $-2.3$ and $0$. The values that 
$\overline{\Delta E_{loc}^{rand}}(i)$ can assume are comprised between 
the two straight lines plotted in Fig.\ref{fig2}a. While the 
dependence of $\overline{\Delta E_{loc}}(i)$
for the random sequence is linear, for a selected sequence
it is quadratic. 
In the case of the selected sequence the quadratic behavior of the 
mutation energy versus the number of closest nearest neighbors induces a 
sharp distinction between bulk sites
($n\gtrsim 3$) and surface sites. As clear form  (Fig. \ref{fig2}a), all
bulk sites are 'hot', while surface sites are 'cold'.
As consequence, 'hot' sites have a certain degree of symmetry 
in target structures, i.e. no one of the sites with the same number of 
nearest neighbors is privileged to the others. The map of the 'hot' and
'cold' sites for a Hamiltonian $\beta=0$ and a 36 monomer target 
structure is presented on (Fig. \ref{fig1}c). 

\subsection{$\beta\not=0$}

If $\beta$ is not zero, but still small in absolute values, it is 
possible to 
minimize the energy of the sequence in two steps, first minimizing 
$H_{\beta=0}=\vec{n}\vec{q}$ and finding
an initial trial optimal sequence $\vec{q}_{tr}$, and then re-minimizing the 
same $H_{\beta=0}$ with an effective 
$\vec{n}_{eff}=\vec{n}+\frac{\beta}{2}\vec{q}_{tr} C$. As was shown in 
Ref. \cite{maksim} this procedure is quite reliable for small $\beta$.

Using the same approximation (\ref{opt}) as in the previous section 
for $\vec{q}_{tr}$ one finds
\begin{eqnarray}
\overline{\Delta E_{loc}^{\beta}}(i) \approx - n_i <q>
(1+\beta <q> \frac{N_i}{n_i}) (n_i - \frac{<q>_{\alpha}}{<q>})
\label{deb}
\end{eqnarray}
where $N_i=\sum_j C_{ij}n_j$ is the number of nearest neighbors of the
nearest neighbors of site $i$ and can assume values in the range
$\{n_i, ..., 4n_i+1\}$. 
The expression of $\overline{\Delta E_{loc}^{\beta}}(i)$,
then, takes into account, through the value of $N_i$, sites lying further 
than
the nearest neighbors of $i$. As consequence of this, we can observe (Fig.
\ref{fig2}b) a broadening of  the
range of values that $\overline{\Delta E_{loc}^{\beta}}(i)$ can assume
for each $n_i$ (we will refer to this broadening as 'energy bands' in our 
further discussion)

From Fig. (\ref{fig2}b)
it follows that the effect of the segregation term in (3) is
to differentiate among the sites with the same number of closest 
nearest neighbors. As $\beta \neq 0$, the second shell of nearest neighbors
starts playing its role, thus introducing a cooperative effects in the
determination of $\overline{\Delta E_{loc}^{\beta}}(i)$. This
splitting of the degeneracies in the $\overline{\Delta E_{loc}^{\beta}}(i)$
at $\beta \not= 0$ can lead to the overlap of 'energy bands'
thus leading to the possibility of encountering a 'hot' site on a surface 
and 'cold' site in the bulk.     
\par
To summarize, from the point of view of single mutations, in the case 
$\beta=0$ the spectrum of mutations is composed by two main parts, namely 
mutations in bulk sites, with high value of $\Delta E_{loc}$, and
mutations in surface sites, with  $\Delta E_{loc}$ close to zero or 
negative. The effect of the coupling term is to broaden the range of the
possible mutation energies for the mutation sites of the same type (bulk 
or surface) mixing the energetic levels
corresponding to mutations in different kind of sites. So, the symmetry of 
hot sites in the target structure can be broken thus allowing some hot 
sites to be found on surface.

\subsection{Explicit Calculations with MJ parameterized Hamiltonian}
To investigate how the 'energy bands' depend upon the strength of the
nonlinear contribution in the interaction matrix we have made some explicit
calculations, using as target structure the $36$mers chain displayed 
on (Fig. \ref{fig1}). 
We have first calculated the best values of $q_\alpha$ and $\beta$
to fit MJ matrix according to Eq. \ref{LTW}, finding $\beta=-0.42$.
Using these 
values for $q_\alpha$, we varied $\beta$ in the range $(-1.5,1.5)$,
optimizing each time the sequence 
with a genetic energy minimization technique. The composition has been
kept fixed 
for all values of $\beta$ and chosen in such a way as to satisfy $q_i 
\sim n_i$ (condition for optimal composition at $\beta=0$).
For the sake of computational convenience, the interaction matrix elements 
have been rescaled to have zero average and  
standard deviation equal to $1.0$. Then, we have plotted the value of 
$\overline{\Delta 
E_{loc}^{\beta}}(i)$ for each lattice site, as function 
of $\beta$. First, we consider the case $\beta < 0$. The raising of 
different 'energy bands' is 
shown in Fig.\ref{fig3}a. For $-0.7 \lesssim \beta < 0$ all the bands
lay into four distinct groups. The first two groups, which correspond to 
the sites with $n_i=1$ or $n_i=2$, contribute to cold sites. The 
other two groups correspond to the sites with $n_i=3$ or $n_i=4$ 
defining the 'warm' and 'hot' sites. This situation is very 
similar to the case of $\beta=0$, where bands do not overlap 
and 'hot' sites are exclusively in the bulk. It is not surprising to find 
the same distribution pattern for the MJ matrix ($\beta=-0.42$)  
(Fig.\ref{fig1}d). 
and HP-like interaction matrix ($\beta=0$) (Fig.\ref{fig1}c).
For $-2.5 \lesssim \beta \lesssim -1.0$ the nonlinear part of the 
interaction matrix starts playing 
its role. As was shown in the previous section the second 
neighbor shell contribute to the value of $\Delta E_{loc}^{\beta}(i)$.
At these values of parameter $\beta$ we observe that some of the energy 
bands corresponding to 'warm' (n=3) and 'cold' sites (n=1,2) mix, 
while the bands corresponding to the 'hot' sites stay well separated 
from the other bands.

In the case of $\beta > 0$ (Fig.\ref{fig3}b) the non-linear effect is 
much more dramatic. For $\beta \gtrsim 0.2$ all bands start mixing 
allowing 'cold' sites penetrate the bulk while pushing 'hot' sites 
on a surface. We can rationalise the different pattern of 'energy 
bands' at $\beta < 0$ and $\beta > 0$ by examining the Eq.\ref{site} and 
Hamiltonian \ref{paramH}. Noticing that $\frac{\beta}{2} \vec{q} C 
\vec{q}$ is of the order of $\frac{\beta}{2} \vec{q} \vec{n} <q>$ we can 
rewrite the Hamiltonian \ref{paramH} in a form 
\begin{equation}
H=\vec{q} \vec{n} (1+\frac{\beta}{2}<q>) + \beta \eta
\end{equation}
where $\eta$ contains the collective effects and 
$\vec{q} \vec{n} (1+\frac{\beta}{2}<q>)$ describes a 'renormalized' 
hydrophobic effect. In the case $\beta<0$, $(\frac{\beta}{2}<q>)>0$ and the 
interaction matrix is largely dominated by the hydrophobic effect, while the 
collective contribution to the Hamiltonian is not strong enough to 
substantially mix the 'energy bands'.

For the $\beta > 0$ case $(\frac{\beta}{2}<q>)<0$ and the renormalized 
hydrophobic effect becomes comparable or smaller than the collective term 
in the Hamiltonian. This allows all bands to mix substantially. 

\subsection{Explicit Calculations for Random Hamiltonian}
Another interesting question which can be addressed is how our conclusions
are modified by the addition of a random term in the $\beta=0$ Hamiltonian.
To study this problem, we have chosen a Hamiltonian in the form 
\begin{equation}
H=\vec{n} \vec{q}+\gamma \frac{1}{2}\sum_{ij}^L \epsilon_{ij}\Delta(r_i-r_j),
\end{equation}
where $\vec{n} \vec{q}$ comes from the parameterization (\ref{LTW}) of the
MJ matrix with $\beta=0$. The virtue of this Hamiltonian is in separation 
in a controllable manner the hydrophobic effect due to the $\vec{n} 
\vec{q}$ and any other non-linear effects are modeled by the random term.
The values of $\epsilon_{ij}$ are taken from a Gaussian distribution
with mean zero and standard deviation $1.0$. In (Fig. \ref{fig4}) 
the 'energy bands' are shown as a function of 'mixing' parameter $\gamma$. 
The overall pattern is exactly the same as discussed in the previous 
section for the case in which non--linear terms in the parameterization 
are switched on leading to the effect of 'band crossing' thus allowing 
'warm' sites to appear on the surface and 'cold sites' in the bulk.
Even in the case of 'band crossing', there is still a clear trend for the 'hot' 
and 'warm' sites to concentrate in the bulk of the structure. This
is due to the fact that bulk sites, building the biggest number of contacts,
still display the strongest response to point mutations.

In the inserts of Figs.\ref{fig3}a,b and Fig.\ref{fig4} shown 
the energy of the optimal non-mutated sequence for different values of 
$\beta$ and $\gamma$, respectively. It is possible to observe a sudden 
decrease in energy of the optimal sequence as $\beta$ and $\gamma$ 
increases signifying that the addition of non-linear terms into a 
Hamiltonian allows for a much better energy minimization.

\section{CONCLUSIONS}

In this work we have considered how the regularity of the interaction
matrix influence the distribution pattern of 'hot' sites. It has been
found that if the matrix is polarized (dominant hydrophobic effect, 
$\beta \sim 0$ 
or $\gamma \sim 0$) this distribution is very simple. All the 'hot' sites can
be found at the positions with maximum number of closest nearest
neighbors (bulk). 

With increasing importance of non-linear terms ($\beta \not=0$,$\gamma 
\not=0$) 
the distribution pattern changes so that the 'hot' and 'warm' sites can
be found in places with smaller number of nearest neighbors 
(surface) while the general trend of the 'hot' sites to fall into a bulk
part of a conformation still holds.    

As pointed out above this can be rationalized by noticing that
if mixing parameter is different than zero each site starts feeling not
only its nearest neighbors but also the more distant sites.
This leads to a collective nature of the interactions giving rise to a
modified distribution pattern of 'hot' sites.

\newpage

\begin{figure}
\caption{
a) Map of the 'hot' (black), 'warm' (dashed) and 'cold' (white) sites for the 
random Gaussian matrix.
b) Map of the mutation sites for the randomly generated Gaussian matrix.
c) Map of the mutation sites for the LTW parameterization of MJ matrix 
with $\beta=0$.
d) Map of the mutation sites for the MJ interaction matrix (HP like model).
\label{fig1}
}
\end{figure}

\begin{figure}
\caption{
a) Energy of mutation as a function of the number of closest nearest 
neighbors for LTW parameterized interaction matrix with $\beta=0$.
$\Delta E_{loc}$ for the optimized sequence exhibits nonlinear behavior
leading to the sharp differentiation of 'hot' and 'cold' sites. While
$\Delta E_{loc}$ for the random sequence can have both positive and 
negative values at any $n$ leading to the possibility of finding a 'hot' 
mutation even at $n=1$.
b) Energy of mutation as a function of the number of closest nearest
neighbors for LTW parameterized interaction matrix with $\beta < 0$.
As second nearest neighbors start contributing to the mutation energy the 
energy line broadens thus allowing for the sites with equal number of 
closest nearest neighbors to have different interaction energies. All 
possible $\Delta E_{loc}$ are confined between the two parabolas shown
on a graph. 
\label{fig2}
}
\end{figure} 

\begin{figure}
\caption{  
Energy bands for 36 mutation sites. The interaction energy matrix is 
based on LTW parameterization with $\beta$ parameter introducing a 
non-linear segregation energy in Hamiltonian. Different line types 
correspond to the sites with different number of closest nearest 
neighbors. So, for example, solid lines correspond to the mutation 
energy of the sites 16 and 27 that have the most number of closest 
nearest neighbors. Sites with 4,3,2 and 1 closest nearest neighbors are 
specified by solid, dashed, solid-dashed and dotted lines respectively. 
a) corresponds to $\beta<0$ b) corresponds to $\beta>0$. In the insert
the energy of the optimal non-mutated sequence is shown for different 
values of $\beta$.
\label{fig3}
}
\end{figure}
 
\begin{figure}
\caption{
Energy bands for 36 mutation sites. The interaction energy matrix is a mix of LTW
parameterized matrix with $\beta=0$ and Gaussian random matrix. The mixing with 
random matrix is controlled by the parameter $\gamma$. At $\gamma=0$ the 
interaction matrix is pure LTW with $\beta=0$ while at $\gamma \sim 2.0$ the 
elements of random matrix become comparable to the elements of the regular matrix.
Different line types correspond to the
sites with different number of closest nearest neighbors. So, for example, 
solid lines correspond to the mutation energy of the sites 16 and 27 that have 
the most number of closest nearest neighbors.
Sites with 4,3,2 and 1 closest nearest neighbors are specified by solid, 
dashed, solid-dashed and dotted lines respectively. In the insert
the energy of the optimal non-mutated sequence is shown for different
values of $\gamma$.  
\label{fig4}
}
\end{figure}

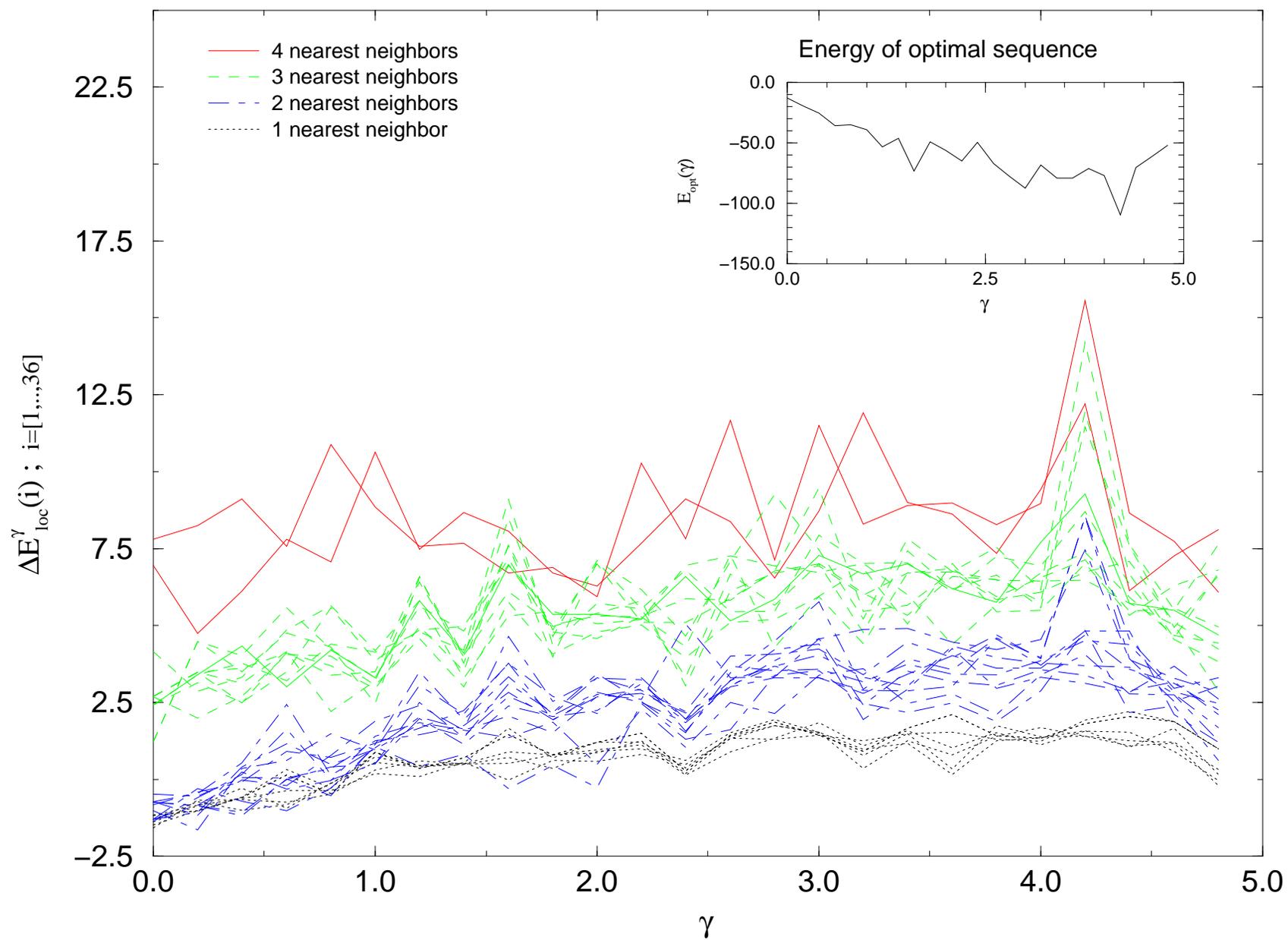
\end{document}